\documentclass[preprint,12pt,sort&compress]{elsarticle}

\usepackage{graphicx}
\usepackage{amssymb}
\usepackage{amsthm}
\usepackage[nodots]{numcompress}
\usepackage{epsfig}
\usepackage{epstopdf}
\usepackage{subfig}
\usepackage{amsmath}
\usepackage{latexsym}
\usepackage{amsbsy}
\usepackage{array}
\usepackage{setspace}
\usepackage{bm}
\usepackage{textcomp}

\journal{}

\begin{document}

\begin{frontmatter}

\title{Ising Spins on Randomly Multi-Branched Husimi Square Lattice: Thermodynamics and Phase Transition in Cross-dimensional Range}

\author[sjtu,zjutz]{Ran ˜Huang \corref{cor1}}
\cortext[cor1]{Correspondence to ranhuang@sjtu.edu.cn}
\address[sjtu]{State Key Laboratory of Microbial Metabolism and School of Life Sciences and Biotechnology, Shanghai Jiao Tong University, Shanghai 200240, China}
\address[zjutz]{Department of Materials Technology and Engineering, Research Institute of Zhejiang University-Taizhou, Taizhou, Zhejiang 318000, China}

\begin{abstract}
An inhomogeneous random recursive lattice is constructed from the multi-branched Husimi square lattice. The number of repeating units connected on one vertex is randomly set to be 2 or 3 with a fixed ratio $P_2$ or $P_3$ with $P_2+P_3=1$. The lattice is designed to describe complex thermodynamic systems with variable coordinating neighbors, e.g. the asymmetric range around the surface of a bulk system. Classical ferromagnetic spin-1 Ising model is solved on the lattice to achieve an annealed solution via the local exact calculation technique. The model exhibits distinct spontaneous magnetization similar to the deterministic system, with however rigorous thermal fluctuations and significant singularities on the entropy behavior around the critical temperature, indicating a complex superheating frustration in the cross-dimensional range induced by the stochasticity. The critical temperature was found to be exponentially correlated to the structural ratio $P$ with the coefficient fitted as 0.53187, while the ground state energy presents linear correlation to $P$, implying a well-defined average property according to the structural ratio. 
\end{abstract}


\end{frontmatter}

\section{Introduction}
The Bethe or Bethe-like recursive lattices generally refer to the fractal arrangements of repeating units recursively connected to neighbors only on the sharing vertex, with no connection bond lies crossing layers. It has become a powerful methodology in various fields such as thermodynamics \cite{1,2}, graph theory\cite{3}, optimization problems \cite{K1,K3} and so on. In statistical physics, one important application of the recursive lattice is to approximate the regular lattice with the identical coordination number to solve a thermodynamic system (e.g. Ising model) on it. As one of few exactly calculable models, it has been proven to be a reliable method \cite{pdg_reliable}, with the advantage of exact calculation and simple iterative approach \cite{exact}, to be applied in numerous physical systems, e.g. alloy \cite{alloy}, spin glass \cite{YoungRev}, polymers \cite{pdg_polymer}, biomacromolecule \cite{Bethe_DNA} etc.

As a versatile extension of the Bethe lattice assembled by single dots and bonds, the Husimi lattice employing simple geographic shapes, such as square, triangle, tetrahedron, hexagon, or cube \cite{Geertsma,Ran_arxiv1,EJ_tetrahedron} has also been developed for decades to describe various systems with particular geographic properties \cite{pdg_Corsi,PDG4,Husimi_RNA}. Similar to the Bethe lattice, The independence feature of units enables the exact calculation on Husimi lattice regardless of the dimensions of the geographic unit, and mean field approximation is unnecessary since the interactions are confined within a unit and not shared by others. The calculation method usually relies on the recursive approach, which is featured as simple and less computation effort costing thanks to the homogeneous self-similar structure \cite{Ran_arxiv1,EJ}.

Nevertheless, the recursive feature is accompanied with several disadvantages of this lattice methodology. Firstly, the repeating structure implies a homogeneous system, it is only suitable to describe systems of uniform texture. Some particular however important cases, e.g. the confined geometry or structural transformation, are enormously difficult, if not impossible, to be simulated by the recursive lattices. Therefore, besides a few investigations on the thermodynamics on the surface/thin film employed moderately inhomogeneous structure to present the boundary of a bulk system\cite{pdg_Chhajer,Ran_arxiv2}, the reports on the application of recursive lattice onto inhomogeneous systems were very rare. Secondly, recursive lattice is considered to be a reliable approximation to regular lattice based on the identical coordination number $q$. Therefore, the manipulation of coordination number(s) is critical in constructing a recursive lattice for particular requirements. While it is easy to draw a regular lattice with an arbitrary $q$, achieving an odd $q$ in recursive lattice usually requires awkward design of unit selection and branch number, and even worse a prime number of $q$ is impossible. Furthermore, when the randomness is necessary in a recursive lattice model, the common method is to add random terms in the Hamiltonian, e.g. a random external field as noise or random exchange couplings parameter $J_{ij}$ \cite{Mezard_cavity,Lage-Castellanos}, while the structural randomness is difficult to be presented due to the homogeneity of recursions.

Therefore, it is expected to be a considerable contribution to this field that if new designs of recursive lattice and calculation methods are developed to handle the above concerns, and make the recursive methodology more versatile in describing inhomogeneous systems. Recently we reported an Husimi lattice of random square-cube recursion, on which the simple Ising model can be solved by conventional exact calculation technique with moderate modification and exhibits well-defined thermal behaviors \cite{JPSJ}. Following the same principle, in this work we have developed a randomly multi-branched Husimi square lattice and solved the simple ferromagnetic Ising model on it. The lattice is featured by randomly two or three square units connected on one vertex, then an inhomogeneous system of variable coordination numbers can be achieved with the identical unit cells. The spontaneous magnetization with critical temperature $T_C$, thermodynamics around the singularity, and the thermal fluctuation caused by stochastic structure, were investigated with the variation of structural ratio.

\section{Modeling and Calculation}

\subsection{Lattice Construction}
\begin{figure}
    \centering{
    	\includegraphics[width=0.5\textwidth]{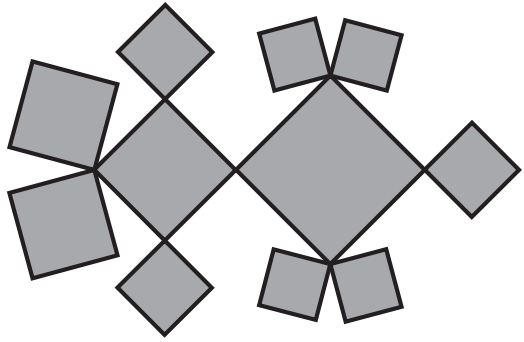}
    	\caption{A demonstration of Husimi lattice with randomly two or three branches joint on one vertex.}
    	\label{fig1}
        }
\end{figure}

The original Husimi lattice was a tree-like graph assembled by squares with two units connecting on one vertex. Since its development, derivative structures of three and more squares connection have been also investigated. In this way, to achieve an inhomogeneous structure with variable $q$, it is a natural choice to have random number of branches connected in the lattice. To keep the investigation simple, in this work we will only study the randomly 2 or 3-branched Husimi square lattice with $q=4$ or $6$ as demonstrated in Fig.\ref{fig1}. A structural ratio $P_2$ or $P_3$ can be defined to indicate the probability to have 2 or 3 branches on one vertex, with obviously $P_2+P_3=1$. However, as a probable reason why this type of lattice has not been reported before, the random structure destroys the recursive homogeneity and then makes the iterative approach unfeasible, therefore the lattice shown in Fig.\ref{fig1} is not the actual model studied in this paper, and particular limitations on the structure must be applied to achieve an exact calculation, which will be detailed later. 

While the branch number is randomly 2 or 3 in this lattice, with the structural ratio $P$ we can define an ``analog branch number" as
\begin{equation}
\text{analog }B=2\times P_2+3\times P_3.
\end{equation}
to present the average branch number of the lattice, and similarly an analog coordination number $q$ can also be defined as $\text{analog }q=2\times\text{analog }B$. By this means, the term ``cross-dimensional" in this paper refers to the gradient of analog $q$: Taking a finite regular cubic lattice with a 2D surface as an example, we have $q=5$ on the surface and $q=6$ in the bulk, then in the near-surface region a randomly sampled site will has a probability to be of either $q$ depending on the depth. Therefore, a gradually variation of analog $q$ from 5 to 6 well represents the cross-dimensional range from the surface to bulk in this case. Similarly, the case of a 2D layer crossing to thin film can be described by the variation of analog $q$ from 4 to 6. For an additional clarification, both 3-branched Husimi square lattice and Husimi cubic lattice have been proved to be a good approximation to the regular cubic lattice \cite{Ran_arxiv1}.

The simplest ferromagnetic spin$\pm 1$ Ising model was applied on the lattice in this paper:
\begin{equation}
E=\underset{<i,j>}{\sum}-J_{ij}S_{i}S_{j},
\end{equation}
without external magnetic field $H$. The weights of one configuration $\gamma$ of a square unit is given by
\begin{equation}
w(\gamma)=exp(-\beta\underset{<i,j>}{\overset{4}{\sum}}-JS_{i}S_{j}),
\end{equation}
where $\beta$ is the reversal temperature as $1/k_{B}T$, the Boltzmann constant $k_{B}$ is set as one. We have the partition function of the entire system as
\begin{equation}
Z=\underset{\Gamma}{\sum}{\prod_{\alpha}}w(\gamma_{\alpha}), \label{PF}%
\end{equation}
where the $\Gamma={\textstyle\bigotimes_{\alpha}}\gamma_{\alpha}$ denotes the state of the lattice as an ensemble of unit $\alpha$.

In this paper we setup a uniform ferromagnetic coupling $J_{ij}=1$, then the state of system only depends on the structural properties. Without external magnetic field, we can expect a half-half probability of spin state on each site at high temperature, a uniform orientation pointing to either up or down of all spins at low temperature, and a spontaneous magnetization occurring in between. The only question being focused on in this paper is that, how this transition behaves in the cross-dimensional situation on an inhomogeneous lattice.

\subsection{Partial Partition Function and Cavity Field}

\begin{figure}
    \centering{
    \subfloat[]{
    \includegraphics[width=0.5\textwidth]{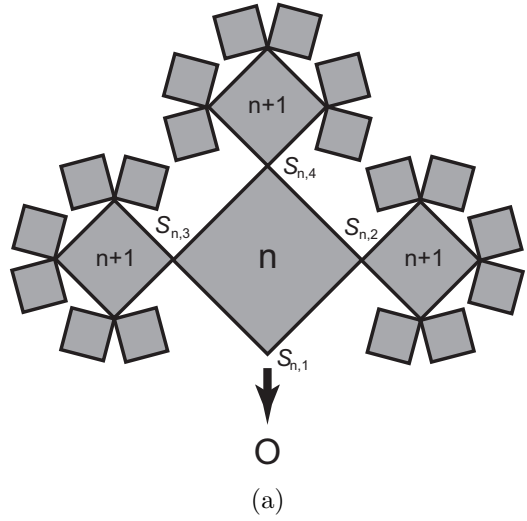}
    }\\
    \subfloat[]{
    \includegraphics[width=0.5\textwidth]{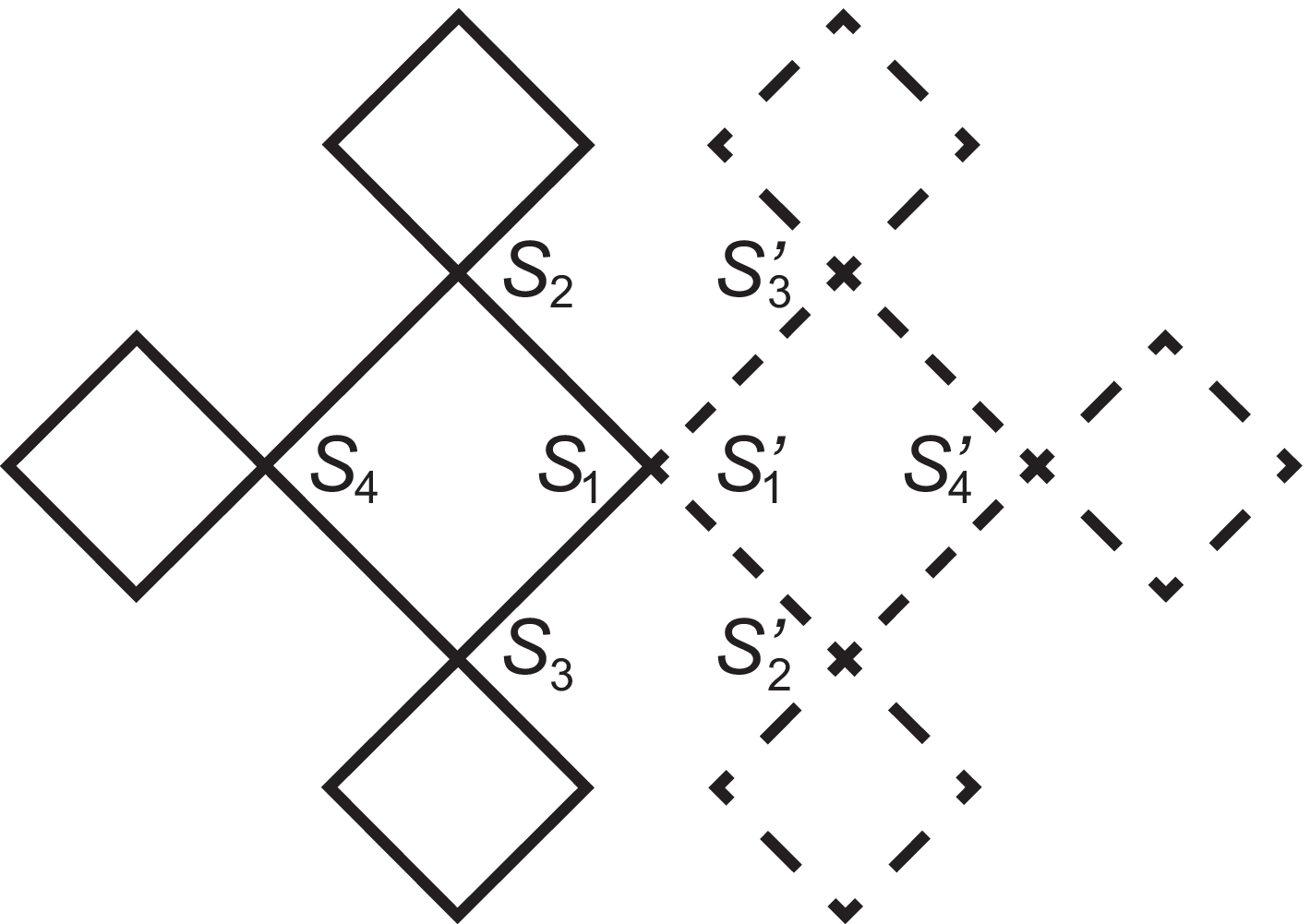}
    }
    \caption{The calculation scheme of the solution $x$ and the free energy. (a) A demonstration of the actual structure employed in this paper to calculate the solutions. To ensure a local homogeneity to facilitate the recursive calculation, branch numbers on the three sites of one square are set to be identical. The recursive calculation towards an imaginary original point $ O $. (b) The calculation of free energy by a replica counterpart to obtain smaller hooked-up lattices to yield the local free energy.}
    
    \label{fig2}
    }
\end{figure}

The lattice is designed of infinite size, nevertheless for an iterative approach it is necessary to imagine an original point where the entire lattice contribute to. Furthermore, the structure must be symmetrical to the original point, and subsequently the symmetry of sub-trees contributing onto one unit is required, otherwise the unique structure of an arbitrary sub-tree is impossible to be tracked and accounted in iterative calculation. Therefore, the unlimited random structure shown in Fig.\ref{fig1} is not the actual lattice we are going to study, and two important principles have to be settled here: 1) the branch number on the vertices of one unit must be the same excluding the base vertex; 2) for any arbitrary square the three sub-trees contributing onto it towards to the original point should be identical. And the branch number on the vertices of different levels are random with the structural ratio $P$. A sample structure is presented in the Fig.\ref{fig2}a. 

Although these two limitations confine a locally ordered configuration on the same levels and impair the randomness of the lattice, from a general view of the infinitely large structure, we still have two or three branched vertex randomly appears with a fixed probability. Therefore,
we may say that this paper discussed a special case of the ideally random multi-branched lattice with identical sub-trees contributions. 

The calculation method basically follows the previous works done by Gujrati's \cite{pdg_Corsi,PDG4,Ran_arxiv1}, here we will briefly derive the method with specific modifications for the random-branched lattice. By defining the partition function of a sub-tree with the base spin fixed as $ +1 $ or $ -1 $ to be the partial partition function (PPF), denoted as $Z_{i}^{2}(\pm1)$, where $i$ is the level of the base spin, at the original site $S_0$ (level $0$) we have the total partition function:

 \begin{equation}
 Z_{0}=Z_{0}^{B}(+1)+Z_{0}^{B}(-1), \label{PF_Origin}
 \end{equation}
The exponent $ B $ is the number of sub-trees (branches) contributing onto $S_{0}$. While in the regular lattice the branch number is a fixed variable, in this work the branch number is a series of random variables on each level, therefore the branch number shall be written as $\{B_i\}$ associated to the level $i$.

As shown in Fig.\ref{fig2}a, for each square unit on level $i$ it always takes in $3(B_{i+1}-1)$ sub-trees contributions from upper level and contribute itself to the unit on the lower level, then by defining $B_i^{\prime}=B_i-1$, for the PPF on the level $n$ we have the recursive relations

 \begin{equation}
Z_{n}(+)=\underset{\gamma=1}{\overset{8}{\sum}}Z_{n+1}^{B_{n+1}^{\prime}%
}(S_{n,2})Z_{n+1}^{B_{n+1}^{\prime}}(S_{n,3})Z_{n+1}^{B_{n+1}^{\prime}}%
(S_{n,4})w(\gamma),\label{PPF+_Recursion}
\end{equation}
\begin{equation}
Z_{n}(-)=\underset{\gamma=9}{\overset{16}{\sum}}Z_{n+1}^{B_{n+1}^{\prime}%
}(S_{n,2})Z_{n+1}^{B_{n+1}^{\prime}}(S_{n,3})Z_{n+1}^{B_{n+1}^{\prime}}%
(S_{n,4})w(\gamma). \label{PPF-_Recursion}
 \end{equation}
With the sites labeling shown in Fig.\ref{fig2}a, on the two sites $S_{n,2}$ and $S_{n,3}$ neighboring to $S_{n,1}$ we have $2B_{n+2}^{\prime}$ sub-trees contributing from level $n+1$, and similar to the $S_{n,4}$ diagonal to $S_{n}$. The $w(\gamma)$ is the local weight of the square confined by the four sites $S_{n,1}$, $S_{n,2}$, $S_{n,3}$, and $S_{n,4}$. In principle this local weight should exclude the weights of base site $S_{n,1}$, $e^{-\beta HS_{s,1}}$, to conduct a cavity field contribution, however this can be ignored since we set the external field $H=0$ in this paper.

With the PPFs defined above, we introduce a ratio $x(S_{n})$ (simplified as $x_{n}$) on the site $S_{n}$

\begin{equation}
x_{n}=\frac{Z_{n}(+)}{Z_{n}(+)+Z_{n}(-)},\label{Ratios}
\end{equation}
as the weights ratio of PPFs with the fixed spin state of $S_{n}$, $x$ denotes the sub-tree contribution to the magnetization field of the site $S_{n}$, i.e. the cavity contribution \cite{cavity}. Therefore this ratio can be treated as the ``solution" of the system to indicate the magnetization of a site. Obviously, the solution of $x=0$ or $1$ indicates a uniform spins orientation to either up or down, and conversely the solution $x=0.5$ implies a half-half probability to have a up or down spin on one site, i.e. the amorphous state without external magnetic field.

Since $x_{n}$ is a function of PPFs on level $n$, and as shown in Eq.\ref{PPF+_Recursion} and \ref{PPF-_Recursion} a PPF is a function of the PPFs on higher level, then $x_{n}$ can be derived as a function of $x$s on higher levels in a recursive fashion:

\begin{equation}
x_{n}=f(x_{n+1}). \label{x_recursion}%
\end{equation}

With an initial seeding input of $ x $, the solution can be obtained by applying the Eq.\ref{x_recursion} many times until reaching a fixed-point solution. Principally, various input seeds should be tried to exhaust all possible fixed points, however this work deals with the simplest ferromagnetic Ising model, and a single solution at a fixed temperature is expected. 

The exact form of Eq.\ref{x_recursion} is derived below: 
Starting from
\begin{align*}
x_{n}=\frac{Z_{n}(+)}{Z_{n}(+)+Z_{n}(-)},\\y_{n}=\frac{Z_{n}(-)}{Z_{n}%
(+)+Z_{n}(-)},
\end{align*}
we define a compact note
\begin{align*}
z_{n}(S_{n})=\left\{
\begin{array}
[c]{c}%
x_{n}\text{ if }S_{n}=+1\\
y_{n}\text{ if }S_{n}=-1
\end{array}
\right.
\end{align*}
In terms of%
\[
A_{n}^{B_n^\prime}=Z_{n}(+)+Z_{n}(-),
\]
we have
\begin{align*}
A_{n}^{B_n^\prime}z_{n}(\pm)  =\sum A_{n+1}^{B_{n+1}^\prime}z_{n+1}^{B_{n+1}^{\prime}%
}(S_{n,2})A_{n+1}^{B_{n+1}^\prime}z_{n+1}^{B_{n+1}^{\prime}}(S_{n,3}%
)A_{n+1}^{B_{n+1}^\prime}z_{n+1}^{B_{n+1}^{\prime}}(S_{n,4})w(\gamma),
\end{align*}
and
\begin{align*}
z_{n}(\pm) =\sum z_{n+1}^{B_{n+1}^{\prime}}(S_{n,2})z_{n+1}^{B_{n+1}^{\prime}}%
(S_{n,3})z_{n+1}^{B_{n+1}^{\prime}}(S_{n,4})w(\gamma)/Q(x_{n+1}),
\end{align*}
where the sum is over $\gamma=1,2,3,\ldots,8$ for $S_{n,1}=+1$, and over
$\gamma=9,10,11,\ldots,16$ for $S_{n,1}=-1$, and where
\[
Q(x_{n+1})\equiv A_{n}^{B_n^\prime}/ A_{n+1}^{3{B_{n+1}^{\prime}}%
};
\]
it is related to the polynomials
\begin{align*}
Q_{+}(x_{n+1}) =\underset{\gamma=1}{\overset{8}{\sum}}%
z_{n+1}^{B_{n+1}^{\prime}}(S_{n,2})z_{n+1}^{B_{n+1}^{\prime}}(S_{n,3}%
)z_{n+1}^{B_{n+1}^{\prime}}(S_{n,4})w(\gamma),\\
Q_{-}(x_{n+1}) =\underset{\gamma=9}{\overset{16}{\sum}}%
z_{n+1}^{B_{n+1}^{\prime}}(S_{n,2})z_{n+1}^{B_{n+1}^{\prime}}(S_{n,3}%
)z_{n+1}^{B_{n+1}^{\prime}}(S_{n,4})w(\gamma),
\end{align*}
according to
\[
Q(x_{n+1})=Q_{+}(x_{n+1})+Q_{-}(y_{n+1}).
\]
In terms of the above polynomials, we can express the recursive relation for
the ratio $x_{n}$ in terms of $x_{n+1}$:
\begin{align*}
x_{n}=\frac{Q_{+}(x_{n+1})}{Q(x_{n+1})}.
\end{align*}

The above process is the static recursive calculation of solutions. For the random multi-branched lattice, on each iteration the program will randomly assign the value of $B$ to be 2 or 3 according to the structural ratio $P_2$, and then execute the calculation for that local level. Therefore, the solutions exhibit a fluctuation around an average value, and the fixed-point solution does not exist. According to our experience, the calculation reaches a ``stable" solutions oscillation $x_{n}\in(\bar{x}\pm \sigma)$ in no more than 2000 iterations. In the program we did the calculation 16,000 times and average the last 11,000 $x$s to reach a reliable $\bar{x}$. In this way, although the exact calculation is executed in each iteration, instead of exact results what we actually obtained is a numerical solution distribution.

\subsection{Calculation of Thermodynamics}

The thermodynamic calculations again follows the same principle detailed in Ref. \cite{Ran_arxiv1,CTP}, while a slight modification is necessary to fit the random conformation case. Herein we firstly review the general method to calculate the free energy on a homogeneous $ B $-branched Husimi lattice.

Imagine we achieved the fixed-point solution in the region around the original point $O$, the $ B $ squares joint on $O$ are indexed as the first level, then there are $3B\times(B-1)$ squares on the level $2$. Now we cut off these sub-trees and hook them up to form $3\times(B-1)$ smaller but identical lattices, the partition function of these lattices are
\[
Z_{2}=Z_{2}^{B}(+)+Z_{2}^{B}(-).
\]
The free energy of the left out local squares is
\[
F_{local}=-T\log  \frac{Z_{1}}{Z_{2}^{3B^{\prime}}}.
\]
We have $4/B$ cites in a square and $B$ squares in the local origin region. The free energy per site is:%
\begin{align*}
F=-\frac{F_{local}}{4}. \label{FreeEnergy/site}%
\end{align*}
By substituting $Z_{n}(+)=A_{n}x_{n}$ and $Z_{n}(-)=A_{n}y_{n},$ we have
\begin{equation}
F=-\frac{1}{4}T\log(\frac{Q^{2B^{\prime}}}{[x^{B}+(1-x)^{B}]^{3B^{\prime}}}). \label{FE}
\end{equation}

An important concern in the inhomogeneous lattice discussed in this paper is that, the original point is merely a conceptual point that indicating the calculation direction, while it cannot be realized as an either 2 or 3 branched point. Therefore, the free energy of a local region should be calculated along the recursive process and averaged, in the same fashion of solutions, to approach the free energy of the whole system.

By this means, an replica counterpart calculation was designed to obtain the free energy along the recursive process, as shown in Fig.\ref{fig2}b. After the solution converged to the stable oscillation, once an iteration of solution was calculated on one site, the immediate calculation of free energy of that site was carried out by hooking an identical replica half-tree onto that site (the shadow part in Fig.\ref{fig2}b). Then as in a complete lattice with the replica counterpart, the central point can be taken as the `origin' and the general method of free energy calculation (Eq.\ref{FE}) can be applied by cutting-off and rejoining the sub-trees on the corresponding sites $S_n$ and $S_n^\prime$ to yield the free energy of the local region. Although this replica part does not exist in the system, due to the perfect symmetry it provides the correct free energy per site of the local region, and the value calculated on the real half is valid. 

We have also tested that, the branch number on the symmetric joint point does not affect the free energy calculation, i.e. hooking one or two imaginary counterparts onto the site to make $ B_i =2 $ or $ 3 $ present the exact same free energy density, i.e. the thermodynamics only depends on the solution $ x $ once it was determined.

\section{Results and Discussion}

\subsection{Spontaneous Magnetization}

\begin{figure}
    \centering{
        \subfloat[]{
    	\includegraphics[width=0.5\textwidth]{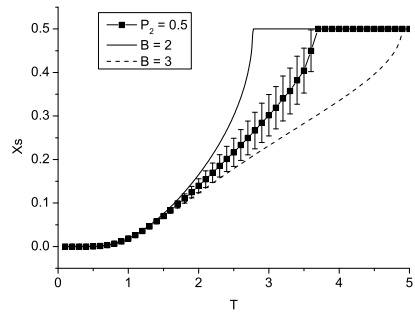}
    	}\\
    	\subfloat[]{
    	\includegraphics[width=0.5\textwidth]{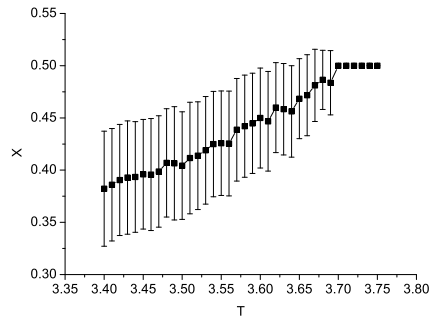}
    	}
    	\caption{(a) The solution $x$ with $P_2=1$, 0.5 and 0. (b) The enlarged $P_2=0.5$ solution curve around the phase transition range.}
       	\label{fig3}
        }
\end{figure}

The average solution $\bar{x}$ with $P_2=0.5$ is presented in Fig.\ref{fig3}a. The solution of deterministic cases of $ B=2  $ and 3 with $P_2=1$ and 0 are also shown for comparison. As expected, the solution of $P_2=0.5$ with its deviation lies inside the area lined out by the two deterministic solutions. Albeit the rigorous fluctuation in the intimidate temperature range, a distinct phase transitions can be figured out with the orientation preference of spins differentiated from $ 0.5 $, i.e. the spontaneous magnetization of simple Ising model.

Figure \ref{fig3}b shows the detailed solution of $P_2=0.5$ around the critical temperature $T_C$. Although the error bar is considerable large merely below $T_C$, the transition is still distinctive. Besides the averaged solution over 11,000 results, we have also investigated several single runs and found that the position of $T_C$ is consistent, i.e. the phase transition of Ising spins is independent of the structural stochasticity. This phenomenon evidenced a well-defined thermodynamics of the randomly multi-branched lattice, and we can be confidential to apply the conventional analysis onto this inhomogeneous lattice.

\subsection{Thermal Behavior and Fluctuation}

\begin{figure}
    \centering{
    \subfloat[]{
	\includegraphics[width=0.5\textwidth]{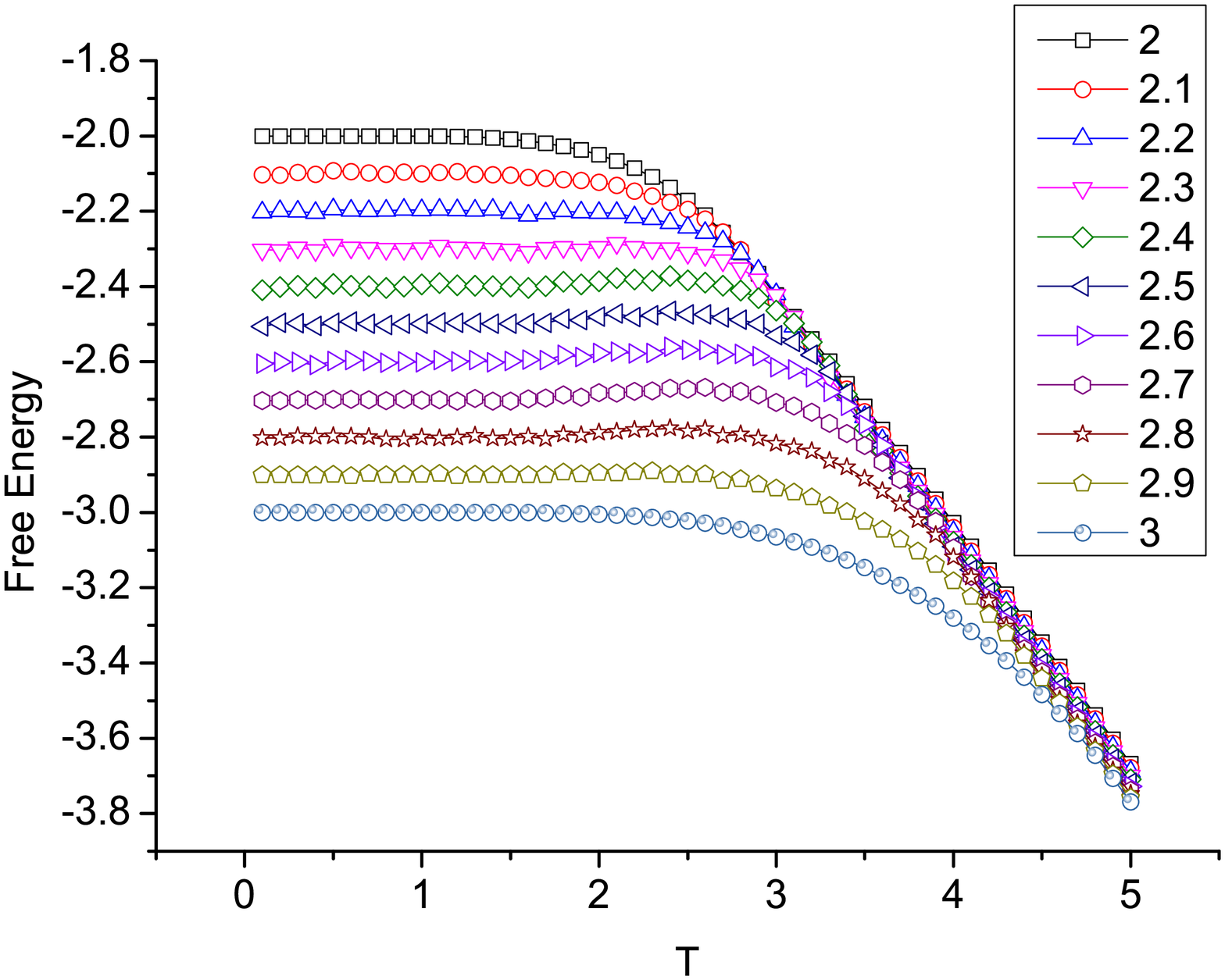}
	}\\
	\subfloat[]{
	\includegraphics[width=0.5\textwidth]{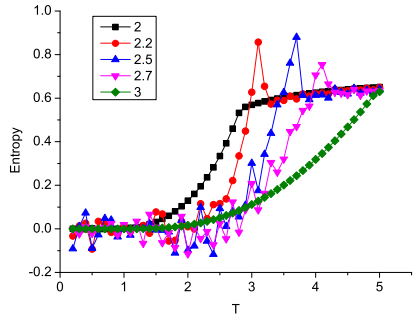}
	}
	\caption{(Color online)(a) The free energy of models with analog $B$ varies from $ 2 $ to $ 3 $. (b) The entropy of models with $B=2$, $2.2$, $2.5$, $2.7$ and $3$. The two deterministic models presents a smooth boundary line, while the stochastic models exhibit rigorous thermal fluctuations in between. Clear singularities can be observed on these curves around the critical temperature, implies the inhomogeneous phases in the cross-dimensional range, where superheating are possible.}

	\label{fig4}
    }
\end{figure}

Ten sets of free energies with $P_2$ varying from 2 to 3 by the increment of 0.1 are shown in Fig.\ref{fig4}a. Similar to the solutions, free energies of cross-dimensional structures lies inside the area outlined by the 2 and 3-branched free energy curves. The uneven curves of the free energies of models with $2<P<3$ imply the thermal fluctuations, which agree with the significant error bar of the solutions presented in Fig.\ref{fig3}.

To more clearly investigate the thermal fluctuation, the entropies derived from $S=-\partial F/\partial T$ of five structures are presented in Fig.\ref{fig4}b. Rigorous fluctuations can be observed on the entropy behaviors of stochastic models, with a superheating extrapolation unphysically surpass the entropy of the amorphous state. These singularities should not be understood as the simple effect of randomness, it actually reveals the inhomogeneity in the cross-dimensional range: For the near surface region of a bulk materials, the layer beneath the surface is at the halfway up the energy landscape, and its state is fragile and easily to be frustrated by the energy flow from surface to bulk or vice versa. At the critical range, the system is in dramatic dynamics, as the surface begins the order-disorder phase transition while the bulk remains ordered, therefore the energy fluctuations is expected to induce locally superheating in between. 
 
\subsection{Branch Ratio vs. Transition Temperature}

\begin{figure}
    \centering{
    \subfloat[]{
	\includegraphics[width=0.5\textwidth]{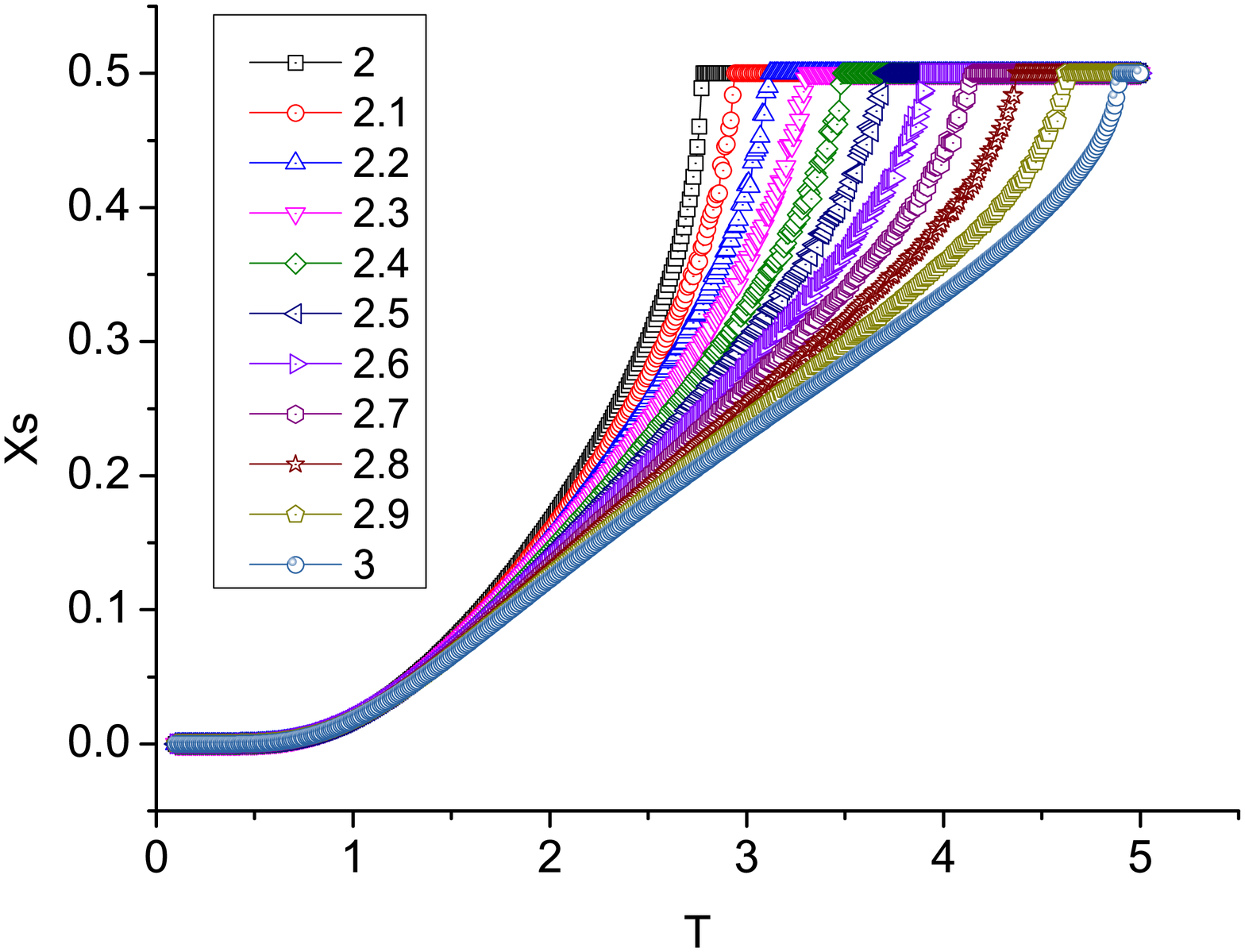}
	}\\
	\subfloat[]{
	\includegraphics[width=0.5\textwidth]{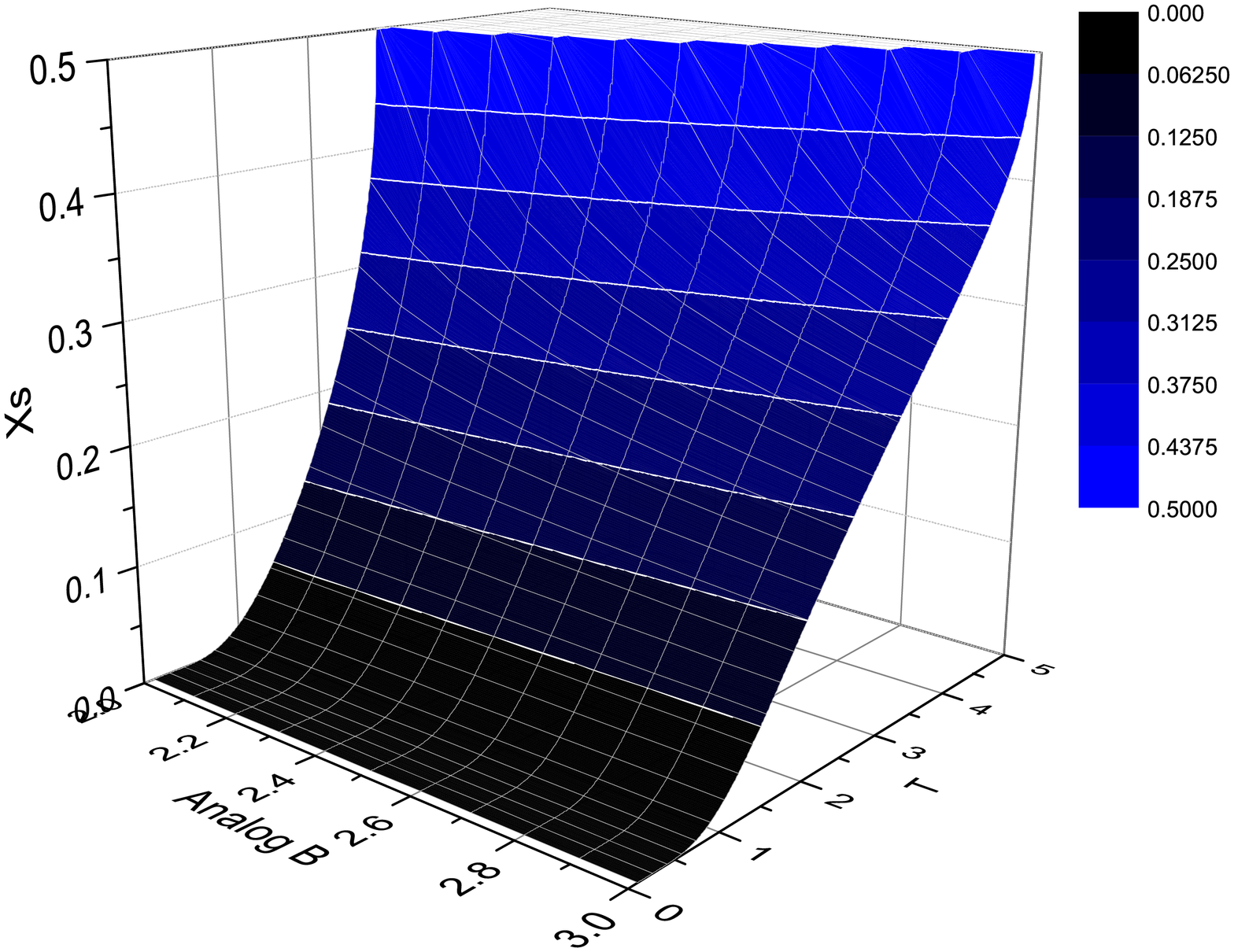}
	}\\
	\subfloat[]{
	\includegraphics[width=0.5\textwidth]{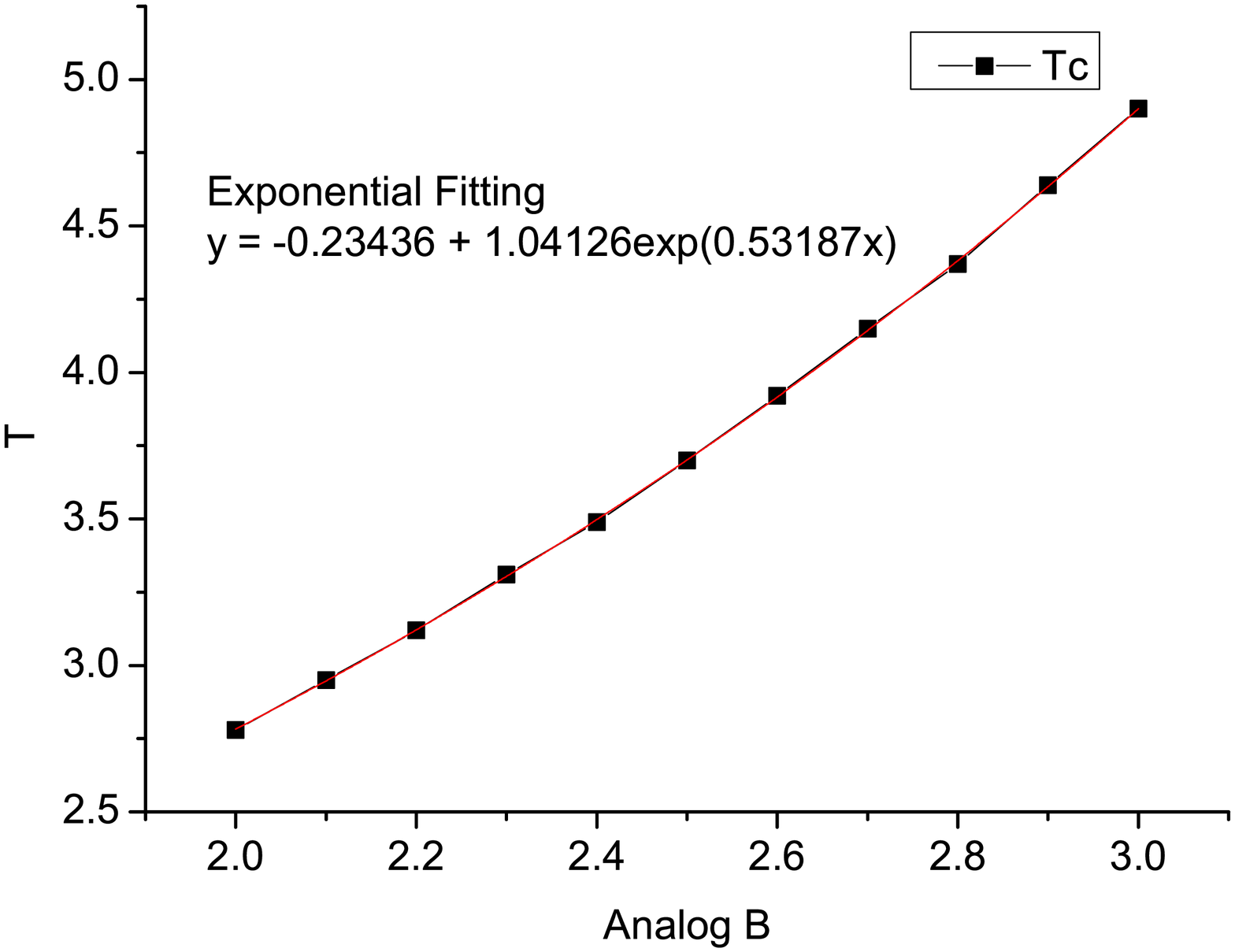}
	}\\
	\caption{(Color online) The transition temperature $T_C$ variation with branch ratio, the ratio is presented by the analog branch number $ B $ by Eq.1: (a) ten sets of solutions of $B$ from 2 to 3 with 0.1 increment; (b) the 3D mapping of the ten sets of solutions in the $T$-$X$-$B$ space; (c) the transition line as the edge of the 0.5 plateau extracted from the 3D mapping.}

	\label{fig5}
    }
\end{figure}

One of the most important aim to design and investigate the inhomogeneous Husimi lattice, is its application in the cross-dimensional range. Figure \ref{fig5}a presents the ten sets of solutions with various analog $B$, and the 3D mapping of these solutions is presented in Fig.\ref{fig5}b. This 3D mapping clearly indicates the gradient in the cross dimensional situation, i.e. the depth from surface to the bulk, or the expansion of a thin film to thick bulk. Herein we can observe from the 3D mapping that it is possible to have different phases on the near-surface region at the same temperature. The edge of the 0.5 plateau is the transition line, which is fitted as
\begin{equation}
T_c=-0.23436+1.04126e^{0.53187B} \label{tranline}
\end{equation}
with $R^2=0.99991$ (Fig.\ref{fig5}c).

Before analyzing the exponential behavior of $T_{\text{c}}$ vs. $B$ presented in Eq.\ref{tranline}, the variation of ground state energy with the analog $B$ should be figured out as a reference. When the systems is at $T=0$, all the spins point to one direction as a perfect crystal and we have the ground state energy $E=F$ with entropy $S=0$. For the homogeneous lattices of $B=2$ and $3$, the ground energy state is simply the number of coupling interactions $E_{T=0,B=2}=4\times\frac{-J}{2}=-2$, or $E_{T=0,B=2}=6\times\frac{-J}{2}=-3$ (one interaction is shared by two spins). For the cross-dimensional lattices in between, the ground state energies $E_{T=0}$ with various analog $B$ exhibit linear behavior as presented in Table\ref{t1}, this phenomenon evidenced that the system has a well-defined average property according to the probability ratio. 

Therefore, the exponential correlation of $T_{\text{c}}$ vs. analog $B$ suggests that the stochasticity might be the single reason to affect the transition behavior near the critical temperature, and we may hypothesize that structural randomness reduces the stability of the system and leads to a lower transition temperature. Nevertheless, in the quantitative aspect, the exponential fitting and the meaning of the parameters in Eq.\ref{tranline} is still unclear. Further investigations on this phenomenon is in progress.

\begin{table}
\caption{Variation of $T_{\text{c}}$ and the ground state energy density with analog $B$.}
\label{t1}
\begin{center}
\begin{tabular}{lll}
\hline
\multicolumn{1}{c}{analog $B$} & \multicolumn{1}{c}{$T_{\text{c}}$} & \multicolumn{1}{c}{$E_{T=0}$}\\
\hline
\multicolumn{1}{c}{$2$} & \multicolumn{1}{c}{$2.78$} & \multicolumn{1}{c}{$-2$}\\

\multicolumn{1}{c}{$2.1$} & \multicolumn{1}{c}{$2.95$} & \multicolumn{1}{c}{$-2.103$}\\

\multicolumn{1}{c}{$2.2$} & \multicolumn{1}{c}{$3.12$} & \multicolumn{1}{c}{$-2.203$}\\

\multicolumn{1}{c}{$2.3$} & \multicolumn{1}{c}{$3.31$} & \multicolumn{1}{c}{$-2.301$}\\

\multicolumn{1}{c}{$2.4$} & \multicolumn{1}{c}{$3.49$} & \multicolumn{1}{c}{$-2.410$}\\

\multicolumn{1}{c}{$2.5$} & \multicolumn{1}{c}{$3.70$} & \multicolumn{1}{c}{$-2.506$}\\

\multicolumn{1}{c}{$2.6$} & \multicolumn{1}{c}{$3.92$} & \multicolumn{1}{c}{$-2.606$}\\

\multicolumn{1}{c}{$2.7$} & \multicolumn{1}{c}{$4.15$} & \multicolumn{1}{c}{$-2.703$}\\

\multicolumn{1}{c}{$2.8$} & \multicolumn{1}{c}{$4.37$} & \multicolumn{1}{c}{$-2.802$} \\

\multicolumn{1}{c}{$2.9$} & \multicolumn{1}{c}{$4.64$} & \multicolumn{1}{c}{$-2.902$} \\

\multicolumn{1}{c}{$3$} & \multicolumn{1}{c}{$4.90$} & \multicolumn{1}{c}{$-3$}\\
\hline

\end{tabular}
\end{center}
\end{table}

\section{Conclusion}

The ferromagnetic Ising model was solved on a randomly multi-branched Husimi square lattice. The lattice was constructed with two or three squares connecting on one with the probability $P_2$ or $P_3$. While the system is deterministic on each end, the structures with variable $P$ in between describes a cross-dimensional situation from 2D to 3D, such as the thin film or the near-surface case. The solution $x$ representing the magnetization of spins was obtained by the cavity partial partition function method, and the thermodynamics such as free energy and entropy were derived from the solutions.

Typical spontaneous magnetization were observed on the random inhomogeneous lattices, the transition occurs on an characteristic $T_C$ regardless of the randomness and thermal fluctuation. The $T_C$, however, behaves an exponential correlation to the structural ratio or analog branch number. Considering the linear relationship of the ground state energy $E_{T=0}$ vs. $B$, the cause of these lowered $T_C$s is hypothesized to be the stochasticity that reduced the stability of the system, while the quantitative meaning of the fitting equation remains unclear. Rigorous fluctuations were observed on the entropy behaviors with singularities of locally supercooling or superheating extrapolation, implying the severe inhomogeneity in the cross-dimensional range around the critical temperature.   

\section{Acknowledgment}

This work is financially supported by the National Natural Science Foundation of China (Grant No. 11505110), the Shanghai Pujiang Talent Program (Grant No. 16PJ1431900), and the China Postdoctoral Science Foundation (Grant No. 2016M591666).

\section{Financial Interests statement}
The author states that there is no competing Financial Interests.

\end{document}